\documentclass{article}

\newtheorem{ThMain}{Theorem}
\renewcommand{\O}{\mathcal{O}}
\newcommand{\E}{\mathcal{E}}
\newcommand{\const}{\hbox{\rm const}}
\newcommand{\Ab}{{\bf{A}}}
\newcommand{\Bb}{{\bf{B}}}

\title
{\bf Asymptotic analysis of a model of autoresonance }
\author
{\bf L.A. Kalyakin
\\
Institute of Mathematics,  Ufa Sci. Centre,  Russian Acad. of
Sci.\\  112,Chernyshevsky str., Ufa,
\\450000, Russia\\
E-mail: klen@imat.rb.ru
\thanks
{The russian version of the paper is presented in "Doklady
Akademii Nauk" ("Doklady Mathematics" in English)} }
\date{March, 27, 2001}

\begin{document}

\maketitle

\begin{abstract}
Autoresonance is a phase locking phenomenon occurring in
nonlinear oscillatory system, which is forced by oscillating
perturbation. Many physical applications of the autoresonance
are known in nonlinear physics. The essence of the phenomenon
is that the nonlinear oscillator selfadjusts to the varying
external conditions so that it remains in resonance with the
driver for a long time. This long time resonance leads to a
strong increase in the response amplitude under weak driving
perturbation. An analytic treatment of a simple mathematical
model is done here by means of asymptotic analysis using a
small driving parameter. The main result is finding threshold
for entering the autoresonance.
\end{abstract}

\section{ Introduction}

We consider a simple mathematical model of forcing
oscillations given by the nonlinear ordinary differential
equation $$u^{\prime\prime}+(1+\gamma u^2)u=2\alpha
f(t)\cos(\varphi(t)), \ t>0;\quad 0<\alpha\ll 1 \eqno (1)$$
where the right hand  side represents a small external force.
The zero initial condition is added here
$$(u,u^{\prime})|_{t=0}=(0,0),\eqno (2)$$ so the system is
starting from stable equilibrium. A nonlinearity may have a
different signs: $\gamma=\pm 1$. The driving amplitude $f$ is
a slow varying function in contrast to the phase function so
that $f^\prime/\varphi^\prime(t)=\o(1),\ \alpha\to 0$. In the
paper we derive the criterion of the autoresonance phenomenon.
Namely, we find a condition under which the system's energy
$\E(t,\alpha)\equiv[(u^\prime)^2+u^2]/2+\gamma u^4/4-u2\alpha
f\cos(\varphi)$ grows up to the order of $\O(1)$ as
$t\to\infty$ while the driver is being small: $0<\alpha\ll
1,\allowbreak\ f(t)=\O(1)$.

Great growth in the energy (and in the swing amplitude as
well) usually takes place due to the resonance in oscillating
system. The simplest model of the resonance phenomenon is
given by the linear equation of harmonic oscillator under
periodic force $$u^{\prime\prime}+\omega^2u=\cos(\nu t).$$
The energy of this system grows infinitely in time, if the
driving frequency coincides with the free frequency
$\omega^2=\nu^2$. In the case of nonlinear system such type of
phenomena is referred to as autoresonance [1-4]. Unlike linear
systems nonlinear systems must have two distinctive features
to be able to enter autoresonance. The first one is that the
driving frequency must be varying, because the free frequency,
depending on the energy, is varying in accordance with the
growth of the oscillating amplitude. The second feature is
that the driving amplitude must exceed a sharp threshold,
depending on the chirp rate of the forcing, as it was noticed
recently [1,2].

An analytic treatment of the threshold phenomenon is done here
by means of asymptotic analysis of the problem (1),(2), using
a small parameter $\alpha$. We believe that an accurate study
of the simple example provides understanding of more
complicated problems. In particular, the fact of the cubic
nonlinearity in (1) does not play any role. Similar results
take place for each oscillating system with a smooth
nonlinearity.

As was pointed out above the free frequency depends on the
energy in nonlinear system. In order to reach large energy of
the swing, using a resonance effect, it is necessary to vary
the driving frequency $d\varphi/dt$ in time, starting from
initial value $d\varphi/dt|_{t=o}=1$. In this way one has to
take into account that energy is small and varying slowly
through initial stage until the amplitude, starting from
zero, increases up to the order of unity. Hence there are two
additional small parameters in the setting of the problem.
The first one is the characteristic frequency mismatch, the
second is the chirp rate, i.e. the rate of change of the
mismatch. We shall identify both small parameters as
$\varepsilon$ and $ \varepsilon^\lambda$ in the phase
function:
$$\varphi=t+\varepsilon^{-\lambda}\Phi(\varepsilon^{1+\lambda}
t),\quad (0<\varepsilon\ll 1,\ \lambda=\const\geq 0),\eqno
(3)$$ In this work the small parameter $\varepsilon$ will be
related to the amplitude parameter $\alpha$ and two types of
the data $f,\varphi(t,\alpha)$ will be considered. Two
different cases are distinguished one from another by the rate
of change of the slow varying functions $f,\varphi(t,\alpha)$
and they are sometimes called as "the rigid frequency
chirping" and "the loose frequency chirping". In each case a
threshold for the driving amplitude is found and entering the
autoresonance is proved in the paper.

We consider here just only the initial stage of amplitude
rising and omit the following stage at which the evolution of
the large amplitude of the order of unity occurs.

Our theory may be considered as an asymptotic analysis of a
small amplitude solution. This approach gives an appropriate
tool in finding the autoresonance phenomena in cases under
consideration. Note, similar results for both problems of the
nonlinear resonance and of the synchronization of oscillation
[5--7] are known. However, they are not directly related to
the autoresonance phenomenon.

\section
{ Anzatz of asymptotic solution} Our approach is a version of
two scale methods, [8--13].  The WKB type anzatz is taken as
an asymptotic solution of the problem (1)--(3):
$$u=\alpha^{1/3}\Big[\Ab(\varepsilon
t,\varepsilon)\exp(i\varphi(t,\varepsilon))+ {\hbox{c.c.}}
+\alpha^{2/3}u_1(t,\varepsilon) +\O(\alpha^{4/3})],\quad
\alpha\to 0.\eqno (4)$$ This is a small amplitude
approximation because of the factor $\alpha^{1/3}$. The
exponent $1/3$ is chosen here in order to take into account
both the nonlinearity and the driver in the slow modulation
of the amplitude $\Ab$. For similar reasons in order to take
into account the frequency mismatch $\varphi^\prime-1$, a
small parameter $\varepsilon$ is related to a driving
parameter: $\varepsilon=\alpha^{2/3}.$ In this way the
nonlinear equation under the zero initial data is derived for
the leading order amplitude
$$2i\Ab^\prime-2\Phi^\prime\Ab+3\gamma|\Ab|^2\Ab=f,\quad
\Ab(\tau)|_{\tau=0}=0;\quad (\tau=\alpha^{2/3}t).\eqno (5)$$ A
modulation of the complex amplitude $\Ab=\Ab(\tau)$ in slow
scale $\tau=\alpha^{2/3}t$ is described by this equation. Our
main discovery is a class of data $\Phi^\prime,f,$ under which
a solution $\Ab(\tau)$ is infinitely increasing as
$\tau\to\infty$. This slow increasing is interpreted as the
initial stage of the autoresonance.

\section
{Rigid driving mode}

The phase function (3) at $\lambda=0$ is considered in the
section. Below this case will be referred to as rigid driving
mode.

It is expedient to represent a complex amplitude $\Ab$ by
means of a pare of real functions $R,\psi$:
$\Ab=R(\tau)\exp(i[\psi(\tau)-\Phi(\tau)])$. One can obtain
explicit solution of the equation (5) for some special data
$f,\Phi$. For example, $$R(\tau)={1\over 2}\int_0^\tau
f(\zeta)\,d\zeta,\quad \psi=\Phi-\pi/2,\quad \Phi(\tau)=
{{3\gamma }\over {2}}\int_0^\tau R^2(\zeta)\,d\zeta.\eqno
(6)$$ So in the case $|f(\tau)|\geq f_0>0$ the amplitude
increases linearly  $|R(\tau)|=\O(\tau),\ \tau\to\infty$ if
the driving phase $\Phi(\tau)$ is related with $f(\tau)$ as
implied by (6). As a result the leading order term of the
asymptotic solution (4) has the order of unity
$u(t,\alpha)=\O(1)$ at time $\tau=\O(\alpha^{-1/3})$, or what
is the same, $t=\O(\alpha^{-1})$.

In general case the existence theorem takes place

{\ThMain Let the functions $f(\tau),\Phi(\tau)$ be continuous
and the $f(\tau)$ be uniformly bounded for all $\tau\geq 0$.
Then there exist a unique solution of the problem (6) for
$\forall\, \tau>0$.}

This result gives a basis for construction of an asymptotics
of the amplitude module $R=R_1\tau+R_0+\O(\tau^{-1/2}),\
\tau\to\infty$. Under assumption that the data have a smooth
asymptotics at infinity:
$$f(\tau)=f_0+f_1\tau^{-1}+\O(\tau^{-2}),\quad
\Phi(\tau)=\Phi_{3}\tau^3+\Phi_{2}\tau^2
+\Phi_1\tau+\Phi_0+\O(\tau^{-1}),\quad\tau\to\infty.\eqno
(7)$$ leading coefficients of an amplitude asymptotics are
calculated: $\gamma R^2_1=2\Phi_3,\ \allowbreak 3\gamma
R_1R_0=2\Phi_2$. So the autoresonance phenomenon, when
$R_1\neq 0$, occurs under the conditions: $$\gamma\Phi_3> 0,\
f_0^2>8|\Phi_3|.\eqno (8)$$ Here the second inequality is just
the threshold condition for entering the autoresonance.

The first correction $u_1$ in (4) is calculated from
linearized equations. There is a secular term in $u_1$, which
increases at infinity $u_1=\O(\tau^3),\ \tau\to\infty$. Hence
the first correction in expansion (4) remains lesser than the
leading term $\alpha |u_1|\ll \alpha^{1/3}|\Ab|$ just only
over a time interval, which is not too long $0<t\ll
\alpha^{-1}$.

The main result in the rigid driving mode case is as follows

{\ThMain Let the right hand side in the equation (1) as given
by (3) under $\lambda=0,\ \varepsilon=\alpha^{2/3},\
(\tau=\alpha^{2/3}t)$ have the properties (7),(8). Then the
system enters autoresonance. The amplitude of the leading
order term in (4) found from equation (5) increases linearly
$|\Ab(\tau)|=\sqrt{2|\Phi_3|}\tau+\O(1),\ \tau\to\infty$.
Asymptotics (4) is available over a time interval, which is
not too long  $0<t\leq\alpha^{-1+\nu},\ \forall \nu>0$.}

\section
{Loose driving mode}

The amplitude equation (5) is nonautonomous and in general
case it can not be solved in the explicit form. However if
both the coefficient $\Phi^\prime(\delta\tau)$ and the right
hand side $f(\delta\tau)$ depend on the slow time
$\delta\tau$ with a small parameter $ 0<\delta\ll 1$ one can
use an asymptotic approach known as adiabatic approximation.
This case will be referred to as loose driving mode below and
will be considered in this section.

We consider the problem (1),(2) with phase driving function
$\varphi$ given by (3) with a small parameter
$\varepsilon=\alpha^{2/3}$ and exponent $\lambda>0$; so the
adiabatic parameter in amplitude equation (5) may be taken
$\delta=\alpha^{2\lambda/3}$. Our main result is an
asymptotics of the amplitude $\Ab=\Ab(\tau,\delta)$ as
$\delta\to 0$, which is valid for large time $\tau\gg
\delta^{-1}$. Using this asymptotics we find the condition
under which the amplitude modulo $|\Ab(\tau,\delta)|$ is
infinitely increasing as $\delta\tau\to\infty$. This slow
increasing is interpreted as the initial stage of the
autoresonance.

Let us go over to the asymptotic constructions. First, the
scaling transformation is performed in the equation (5)
$$\Ab(\tau,\delta)=(f/3\gamma)^{1/3}\Bb(\zeta, \delta), \quad
\zeta=\eta/\delta,\quad \eta= {{(3\gamma)^{1/3}}\over
{2}}\int_0^{\delta\tau} f^{2/3}(\xi)\,d\xi.\eqno (9)$$ Then
the problem for the complex amplitude $\Bb=P+iQ$ is
represented by a system of differential equations for two
functions, namely for real and imaginary parts
$$Q^\prime-[P^2+Q^2-\Omega(\eta)]P=-1+\delta F(\eta)Q,\quad
Q(\zeta,\delta)|_{\zeta=0}=0,\eqno (10)$$
$$P^\prime+[P^2+Q^2-\Omega(\eta)]Q=\delta F(\eta) P,\quad
P(\zeta,\delta)|_{\zeta=0}=0.\eqno (11)$$ Coefficients
$$\Omega(\eta)= 2\Phi^\prime/(3\gamma)^{1/3}
f^{2/3}(\delta\tau),\ \allowbreak F(\eta)= -2f^\prime /
(3\gamma)^{1/3}f^{5/3}(\delta\tau)$$ depend on the new slow
time $\eta=\delta\zeta,\ 0<\delta\ll 1 $. These equations are
considered either on the semiaxis $\zeta>0$ or $\zeta<0$
depending upon the signs of nonlinearity $\gamma=\pm 1$.

A small parameter $\delta$ occurs as the factor in equations
(10),(11) and it defines the slow time in data as well. To
solve such perturbed problem we apply an asymptotic method,
which is ascribed to Bogolubov, Krylov, Kuzmak, Haberman,
[8,11,12]. The basis of the approach is a two-parametric
periodic solution of the unperturbed problem under frozen data
$\Omega\equiv\const$. In the case under consideration the
unperturbed system (as $\delta=0,\ \Omega\equiv\const$) is
autonomous and hamiltonian: $$Q^\prime-[P^2+Q^2-\Omega]P=-1,$$
$$P^\prime+[P^2+Q^2-\Omega]Q=0.$$ The system is integrable
and an explicit integral representation of the solution can
be obtained. But there is no need in an explicit formula to
grasp the situation and to find the condition for entering
the autoresonance. To this end the Hamiltonian as the first
integral may be used $${1\over 4}
(P^2+Q^2-\Omega)^2-P=\const.$$ One can see the family of
periodic solutions from the picture of the phase space
portrait, where nearly all trajectories are closed except for
two embedded separatrixs. To parametrize the solution it is
reasonable to use both the plane area $\Pi$ covered by
trajectory and the phase shift $s$ in the form
$P,Q=P_0,Q_0(\zeta+s,\Pi,\Omega)$. Moreover the solution
depends on the parameter $\Omega$ since $\Omega$ is present
in the equations.

The behavior of separatrixs depending upon the $\Omega$ has
crucial role in detection of the autoresonance. Both these
lines tend to the large circle $P^2+Q^2=\Omega$ as
$\Omega\to\infty$, while the distance between them is small
and has the order $\O(\Omega^{-1/4})$ outside some
neighbourhood of the saddle point. However, the area between
separatrix loops increases and it is minorized by a large
magnitude $m\Omega^{1/4},\ (m=\const>0)$, since the
separatrix length has the order of $\O(\Omega^{1/2})$. In the
considered case with frozen data the solution $P_0,Q_0$
starting from zero is periodic and finite at any time
$\zeta$. Its phase space trajectory depends on the parameter
$\Omega$. It may be located either inside the inner
separatrix or between them.

We return to the perturbed problem (10),(11) in which either
$\Omega(\eta)\neq\const$ and $\delta F(\eta)\neq 0$ or only
one of the inequalities $\Omega(\eta)\neq\const,\ \delta
F(\eta)\neq 0$ holds. One can construct an asymptotic solution
which is valid over a long time interval by using both the
adiabaticity of the coefficient $\Omega$ and the smallness of
the factor $\delta$. Following [7-9] we introduce the
two-scale representation of the solution. In this approach
the leading order term is taken as the unperturbed solution
with additional dependence on the slow time
$$Q_0=Q_0(\sigma,\Pi(\eta),\Omega(\eta)),\
P_0=P_0(\sigma,\Pi(\eta),\Omega(\eta))$$ while the fast
variable is $\sigma=\delta^{-1}S_1(\eta)+S_0(\eta)$. So an
asymptotic solution of (10),(11) is represented by a pare of
functions which are periodic with respect to fast variable
$\sigma$ and depend on slow time by means of
$\Pi,\Omega(\eta)$. In this approach it is necessary to find
slow varying functions $\Pi,S_1,S_0(\eta)$ in order to
identify the leading order term of the asymptotic solution .

We are interested in the solution in which the amplitude
$|P|+|Q|$ is an increasing function of the slow time $\eta$.
Evidently, the structure of the fast variable $\sigma$ is not
needed at this stage. The main result in the section is based
on finding the slow varying function $\Pi(\eta)$. It is
obtained from differential equation, [11,12] which turns up
trivial due to the appropriate parametrization
$$\partial_\vartheta\Pi=F(\vartheta)\Pi,\quad
\Pi(\vartheta)|_{\vartheta=0}=\Pi_0.$$ Here $\Pi_0$ is the
area covered by the initial trajectory (as $\eta=0$). In this
way a remarkably simple formula is obtained:
$$\Pi(\eta)=\Pi_0|f(0)/f(\delta\tau)|.$$ In the special case
when $f\equiv\const$ the area is an adiabatic invariant
$\Pi\equiv\const$.

Thus using (6), one can write down an adiabatic approximation
of the solution of the amplitude problem (5) as follows
$$\Ab(\tau,\delta)=(f(\delta\tau)/3\gamma)^{1/3}
[P_0(\sigma,\Pi(\eta),\Omega(\eta))+
iQ_0(\sigma,\Pi(\eta),\Omega(\eta))]+\tilde\Ab(\tau,\delta),
\eqno (12) $$ Here the remainder  $\tilde\Ab(\tau,\delta)\to 0
\ {\hbox{ as}} \ \delta\to 0.$ As to the remainder it is known
[13] that it is evaluated through small parameter
$\tilde\Ab(\tau,\delta)=\O(\delta),\ \delta\to 0$ and the
estimate is uniform over long time interval
$0\leq\zeta\leq\eta_0\delta^{-1},\ (\eta_0=\const>0)$. But
this strict result is not sufficient for us, because the
autoresonance phenomenon is detected only at very far time
$\zeta\gg \delta^{-1}$ where the leading term of the
asymptotics (12) may be increasing. There is no conflict
since the remainder is small $\tilde\Ab(\tau,\delta)=\o(1),\
\delta\to 0$ over a longer interval, on which the slow time
can become large $\eta=\delta\zeta\gg 1$. However, one has to
keep in mind that the remainder becomes worse with the growth
of time. Hence the approximation given by (12) is valid so
far as the remainder can be neglected with respect to leading
term. Analysis of the first correction in the adiabatic
asymptotic may suggest a limiting time interval but we do not
treat corrections in this work. One can say a priori that an
adiabatic approximation of the amplitude is valid until the
small amplitude expansion (4) is suitable.

The main result of the section is derived from the formula
(12) under the relation $\delta=\alpha^\mu,\
(\mu=2\lambda/3)$.

{\ThMain Let the right hand  side in the equation (1) have
both the amplitude and the phase $f=f(\theta),\
\varphi=t+\alpha^{-\mu}\Phi(\theta),\
(\theta=\alpha^{\mu+2/3}t,\ \mu>0)$ which satisfy
$|f(\theta)|\geq|f(0)|;\ |\Phi^\prime(\theta)|\to \infty$ as $
\theta\to\infty$; let the function $
\Phi^\prime/\gamma^{1/3}f^{2/3}(\theta)\to\infty$ be
monotonous and increase to infinity  of $\theta$. Then
entering the autoresonance depends on the initial value of
the parameter $\Omega(0)=2\Phi^\prime(0)/(3\gamma)^{1/3}
f^{2/3}(0).$ Autoresonance does not occurs if
$\Omega(0)>3/2^{1/3}$. Autoresonance occurs under
$\Omega(0)<3/2^{1/3}$. In the last case the leading order
term of an amplitude asymptotics is determined by the driving
frequency as follows $$|\Ab(\tau,\alpha^\mu)|= \Big({2\over
3}{|\Phi^\prime(\theta)|\Big)^{1/2}}+
\O(|\Phi^\prime|^{-1/4})+\O(\alpha^\mu), \ \alpha\to 0,\
\theta\to\infty.$$}

In order to prove the theorem it is enough to observe a slow
deformation of the phase plane trajectory
$P_0,Q_0(\sigma,\Pi(\eta),\Omega(\eta))$ under variation of
the parameter $\eta$. In the first case the trajectory is
located inside loop of the inner separatrix and it encircles
the equilibrium point near zero. In the second case the
trajectory is located between separatrix loops and it
encircles another equilibrium point near $P=\sqrt
{\Omega(\eta)}$. Under given conditions the plane area $\Pi$
covered by trajectory is not increasing, while separatrix
loops grows like the circle $P^2+Q^2=\Omega$ as
$\Omega\to\infty$. Hence in the first case the trajectory
remains in bounded part of the phase plane for all time. In
the second case one has to take into account that the plane
area between loops is increasing. Hence the trajectory remains
between loops at any time and it is going to infinity as
$\Omega\to\infty$.

\section
{Acknowledgments}

This work may be considered as an attempt of a mathematician
to understand physicist's papers [1-4]. The author is
grateful Dr. A.~Shagalov for suggestion of the subject. This
research has been supported by the Russian Foundation of the
Fundamental Research under Grants 00-01-00663, 00-15-96038
and by INTAS under Grant 99--1068


\end{document}